**Original Paper**

# Using smartphones and wearable devices to monitor behavioural changes during COVID-19

Shaoxiong Sun[1], PhD; Amos A Folarin[1,2], PhD; Yatharth Ranjan[1], MSc; Zulqarnain Rashid[1], PhD; Pauline Conde[1], BSc; Callum Stewart[1], MSc; Nicholas Cummins[3], PhD; Faith Matcham[4], PhD; Gloria Dalla Costa[5], MD; Sara Simblett[4,13], PhD, Letizia Leocani[5], MD, PhD; Per Soelberg Sørensen[6], MD; Mathias Buron[6], MD; Ana Isabel Guerrero[7], MSc; Ana Zabalza[7], MD; Brenda WJH Penninx[8], PhD; Femke Lamers[8], PhD; Sara Siddi[9], PhD; Josep Maria Haro[9], MD; Inez Myin-Germeys[10], PhD; Aki Rintala[10], MSc; Til Wykes[4,13], PhD, Vaibhav A Narayan[11], PhD; Giancarlo Comi[12], MD; Matthew Hotopf[4,13], PhD; Richard JB Dobson[1,2], PhD; RADAR-CNS consortium[14]

[1]The Department of Biostatistics and Health informatics, Institute of Psychiatry, Psychology and Neuroscience, King's College London, London, UK
[2]Institute of Health Informatics, University College London, London, UK
[3]Chair of Embedded Intelligence for Health Care & Wellbeing, University of Augsburg, Germany
[4]Institute of Psychiatry, Psychology and Neuroscience, King's College London, London, UK
[5]University Vita Salute San Raffaele, Neurorehabilitation Unit and Institute of Experimental Neurology, IRCCS Ospedale San Raffaele, Milan, Italy
[6]Danish Multiple Sclerosis Centre, Department of Neurology, Copenhagen University Hospital Rigshospitalet, Copenhagen, Denmark
[7]Multiple Sclerosis Centre of Catalonia (Cemcat), Department of Neurology/Neuroimmunology, Hospital Universitari Vall d'Hebron, Universitat Autònoma de Barcelona, Barcelona, Spain
[8]Department of Psychiatry, Amsterdam UMC, Vrije Universiteit and GGZinGeest, Amsterdam, The Netherlands
[9]Parc Sanitari Sant Joan de Déu, CIBERSAM, Universitat de Barcelona, Sant Boi de Llobregat, Barcelona, Spain
[10]Centre for Contextual Psychiatry, Department of Neurosciences, KU Leuven, Leuven, Belgium
[11]Janssen Research and Development LLC, Titusville, NJ, USA
[12]Institute of Experimental Neurology, IRCCS Ospedale San Raffaele, Milan, Italy
[13]South London and Maudsley NHS Foundation Trust, London, UK
[14]www.radar-cns.org

**Corresponding author**
Shaoxiong Sun
Email: shaoxiong.sun@kcl.ac.uk
Telephone: +44 (0)20 7848 0951
SGDP Centre, IoPPN
King's College London
Box PO 80
De Crespigny Park, Denmark Hill
London
SE5 8AF

# Abstract

**Background:** In the absence of a vaccine or highly effective treatment for COVID-19, countries have adopted Non-Pharmaceutical Interventions (NPIs) such as social distancing and full lockdown. An objective and quantitative means of passively monitoring the impact and response of these interventions at a local level is urgently required.

**Objective:** We aimed to explore the utility of the recently developed open-source mobile health platform RADAR-base as a toolbox to rapidly test the effect and response to NPIs aimed at limiting the spread of COVID-19.

**Methods:** We analysed data extracted from smartphone and wearable devices and managed by the RADAR-base from 1062 participants recruited in Italy, Spain, Denmark, the UK, and the Netherlands. We derived nine features on a daily basis including time spent at home, maximum distance travelled from home, maximum number of Bluetooth-enabled nearby devices (as a proxy for physical distancing), step count, average heart rate, sleep duration, bedtime, phone unlock duration, and social app use duration. We performed Kruskal-Wallis tests followed by post-hoc Dunn's tests to assess differences in these features among baseline, pre-, and during-lockdown periods. We also studied behavioural differences by age, gender, body mass index (BMI), and educational background.

**Results:** We were able to quantify expected changes in time spent at home, distance travelled, and the number of nearby Bluetooth-enabled devices between pre- and during-lockdown periods ($P < .001$ for all five countries). We saw reduced sociality as measured through mobility features, and increased virtual sociality through phone usage. People were more active on their phones ($P < .001$ for Italy, Spain, and the UK), spending more time using social media apps ($P < .001$ for Italy, Spain, the UK, and the Netherlands), particularly around major news events. Furthermore, participants had lower heart rate ($P < .001$ for Italy, Spain; $P = .02$ for Denmark), went to bed later ($P < .001$ for Italy, Spain, the UK, and the Netherlands), and slept more ($P < .001$ for Italy, Spain, and the UK). We also found that young people had longer homestay than older people during lockdown and fewer daily steps. Although there was no significant difference between the high and low BMI groups in time spent at home, the low BMI group walked more.

**Conclusions:** RADAR-base, a freely deployable data collection platform leveraging data from wearables and mobile technologies, can be used to rapidly quantify and provide a holistic view of behavioural changes in response to public health interventions as a result of infectious outbreaks such as COVID-19. RADAR-base may be a viable approach to implementing an early warning system for passively assessing the local compliance to interventions in epi/pandemics and could be particularly vital in helping ease out of lockdown.

# Introduction

On 11 March 2020, the World Health Organisation (WHO) declared the rapidly spreading SARS-CoV-2 virus outbreak a pandemic. This novel coronavirus is the cause of a contagious acute respiratory disease (COVID-19), which was first reported in Wuhan,

Hubei Province, China [1–3]. As of 1 July 2020, it had infected over ten million people and spread to 213 countries and territories around the world [4]. While precise statistics on mortality are being determined, COVID-19 can be deadly with an estimated 1% case fatality rate, and this rate increases dramatically for the elderly and vulnerable who have underlying health problems [5,6]. The outbreak of COVID-19 has placed an unprecedented burden on healthcare systems in most-affected countries and has resulted in considerable economic losses and possible deep global recession [7,8].

To date, there is no vaccine or highly effective treatment. The widely adopted strategy has been the use of Non-Pharmaceutical Interventions (NPIs) such as social distancing and even full lockdown in order to control the spread of the virus and ease pressure on health and care systems [9,10]. NPIs have been implemented in many countries including China, Italy, Spain, the United Kingdom (UK), and the Netherlands. These measures have been shown to considerably reduce the new confirmed cases [9]. Key to the success of NPIs is the timing of these interventions and the response of the population, both of which might differ among countries, and could necessitate further interventions in the case of low compliance either nationally or locally. Furthermore, eleven trillion dollars of fiscal measures have been announced by more than two-thirds of governments across the world in an attempt to mitigate the fallout from the pandemic and lockdown [11]. Therefore, we urgently require an objective and quantitative way to monitor population behaviour to assess the impact and response of such interventions. Additionally, we need to monitor for the potential effects of a rebound in cases in the winter months as social distancing measures are relaxed in order to strategise and understand where course corrections are required. Similarly, understanding potential seasonal forcing of COVID-19 will require a good understanding of the effects of different NPIs so they can be factored out.

The increasing availability of wide-bandwidth mobile networks, smartphones, and wearable sensors makes it possible to collect near-real-time high-resolution datasets from large numbers of participants and greatly facilitates remote monitoring of behavior [12–14]. By leveraging sensor modalities in smartphones which includes Network/Global Positioning System (GPS) location tracking, and Fitbit devices which includes step counts and heart rate, it is possible to access mobility and even wellness for the population. To manage the data collected from multiple sensor modalities and mobile devices, platforms such as the open-source RADAR-base (radar-base.org) mobile health platform have been developed [15]. This platform has been used to enable remote monitoring in a range of use cases including central nervous system diseases (Major Depressive Disorder (MDD), epilepsy and Multiple Sclerosis (MS)) as part of the IMI2 RADAR-CNS major programme (radar-cns.org) [16].

In this paper, we explored the utility of the RADAR-base platform as a toolbox to test the effect and response of NPIs aimed at limiting the spread of infectious diseases such as COVID-19 by leveraging participant data already collected from November 2017 onwards as part of the ongoing RADAR-CNS studies [15–17]. Specifically, we created measures of mobility (as a proxy of physical distancing), phone usage (as a proxy of virtual sociality), and functional measures (heart rate and sleep), and compared these features among the baseline, pre- and during-lockdown periods. Furthermore, we also provided a joint analysis of these features to provide a holistic view of and interpret these behavioral changes during COVID-19.

## Methods

### Data collection

The RADAR-CNS studies were approved by all local ethics committees, and all participants signed informed consent [17]. We included 1062 participants recruited in five European countries: Italy, Spain, Denmark, the UK, and the Netherlands. The data have been collected for the purpose of finding new ways of monitoring MDD (Spain (150), the Netherlands (103) and the UK (316)) and MS (Milan, Italy (208); Barcelona, Spain (179); and Copenhagen, Denmark (106)) using wearable devices and smartphone technology to improve patients' Quality of Life (QoL), and potentially to change the treatment of these and other chronic disorders. As we focused on country-level behavioural changes in response to the NPIs, we aggregated data collected in Spain and did not focus on analysing differences between participants with MDD and MS. Passive participant data, that is data that did not require conscious participant engagement, were collected continuously on a 24/7 basis through a smartphone and a Fitbit device, which included location, Bluetooth, activity, sleep, heart rate and phone usage data. In this study, we used participants' own Android smartphones where available and provided a participant with a Motorola G5, G6, or G7 if participants had an iPhone or did not have a smartphone. For Fitbit devices, Fitbit Charge 2 devices were given to participants and then Fitbit Charge 3 devices were given to the recently recruited participants when Fitbit Charge 2 devices were no longer available. We asked participants to wear the device on their non-dominant hand. Although not used for this study, active data were also collected, which required clinicians or participants to fill out emailed surveys (e.g. Inventory of Depressive Symptomatology (Self-Report)), app-delivered questionnaires (e.g. Patient Health Questionnaire), or perform short clinical tests (e.g. Expanded Disability Status Scale).

The data collection and management were handled by the open-source mHealth platform RADAR-Base [15]. The platform provides high scalability, interoperability, flexibility and reliability while allowing the freedom for anyone to deploy. Due to the streaming first nature of the platform, it is also easy to aggregate, analyse, and provide insights into the data in real-time, hence making the results of this work potentially deployable for localised monitoring and targeted interventions.

### Feature extraction

To study physical-behavioural changes in response to COVID-19 NPIs, we examined participants' mobility by analysing relative location and Bluetooth data from smartphones and step count data from Fitbit devices. We investigated phone unlock duration and social app use duration to study online social-behavioural changes. Functional measures such as sleep and heart rate from Fitbit devices were also analysed to identify possible changes as a result of lockdown. A full list of features is presented in Table 1. These features were extracted for each participant every day. The daily features were calculated using the data from 6 am on the day to 6 am on the next day for all features except total sleep duration and bedtime, where 8 pm was used as the starting time point and 11 am the finishing. When no data were found in a data modality for a participant on a day due to the participant not wearing the Fitbit device or not using the smartphone, we did not calculate the feature derived from that data modality on that day.

**Table 1.** A full list of extracted features

| Category | Modality | Features | Extraction |
| --- | --- | --- | --- |
| Mobility | Smartphone location | Homestay | The time spent within 200m radius of home location (determined using DBSCAN) |
| | | Maximum travelled distance from home | The maximum distance travelled from home location |
| | Smartphone Bluetooth | Maximum number of nearby devices | The maximum number of Bluetooth-enabled nearby devices |
| | Fitbit step count | Step count | Daily total of Fitbit step count |
| Functional measures | Fitbit sleep | Sleep duration | Daily total duration of sleep categories (light, deep, and rem) |
| | | Bedtime | The first sleep category at night |
| | Fitbit heart rate | Average heart rate | The daily average heart rate |
| Phone usage | Smartphone user interaction | Unlock duration | The total duration of phone in the unlocked state |
| | Smartphone usage event | Social app use duration | The total duration spent on social apps (Google Play categories of Social, Communication, and Dating) |

The smartphone-derived location data were sampled once every five minutes by default, with longer sampling durations dependent on network connectivity. Spurious location coordinates were identified and removed if they differed from preceding and following coordinates by more than five degrees. Home location was determined daily by clustering location data between 8 pm and 4 am with the mean coordinate of the cluster that the last coordinate belonged to being used. This choice was made because the largest cluster may not be the home location for a single night but the last location before phones shut down had a higher probability to be home location for that night. The clustering was implemented using Density-Based Spatial Clustering of Applications with Noise (DBSCAN) [18]. A duration gated by two adjacent coordinates was regarded as a valid homestay duration on the condition that both coordinates were no further than 200 meters from the home location. A duration longer than one hour was excluded due to the large proportion of missing data when compared to the five-minute sampling duration. All valid home stay durations between 8 am and 11 pm were summed to calculate daily homestay. Daily maximum distance from home was also computed based on the coordinates in the same period.

Bluetooth data, including the number of nearby and paired devices, were also collected from smartphones, which were sampled every hour. The daily maximum number of nearby devices was used as a mobility feature. An increased number of nearby devices (typically other phones) detected may indicate more other users' presence in the vicinity, which therefore can serve as a proxy of physical distancing.

In addition to mobility features extracted from smartphones, daily step count was taken from the Fitbit device, which was computed as the total steps a participant walked every day. Likewise, daily sleep duration was computed as the summation of three Fitbit-output sleep categories (light, deep, and rem) sampled every 30 seconds for the time interval from 8 pm to 11 am the next day. Bedtime was defined as the time of the first sleep category reported by Fitbit after 8 pm. Note that the sleep categories referred to the sleep stages provided by Fitbit API [19], which are not equivalent to the medical sleep stages. Finally, daily mean heart rate was calculated daily by averaging the Fitbit-output heart rate readings, sampled every five seconds at best. This sampling interval may be longer depending on Fitbit proprietary algorithms for quality scoring and network connectivity.

To explore changes in phone usage, daily unlock duration was calculated by summing time periods starting with the unlocked state and ending with the standby state. Single intervals longer than four hours were excluded, which might result from a missing standby state or unintentionally leaving the phone unlocked. App usage was quantified by classifying apps according to categories listed on Google Play [20]. As we were particularly interested in cyber social interactions, we focused on the daily use time of social apps, including the Google Play categories of Social, Communication, and Dating. Among them are Facebook, Instagram, and WhatsApp.

### Data analysis

We plotted how the features evolved over 1.5 years for each country investigated. The participants' daily median, 25th percentile, and 75th percentile of each feature were calculated and then plotted. A minimum of 20 participants' data points was a prerequisite for calculation for any given day to reduce variance and noise. To facilitate interpretation, we also marked time points of public announcements related to lockdown policies [21].

To examine changes in mobility, functional measures, and phone usage induced by the lockdowns, comparisons among baseline, pre-, and during-lockdown on the daily median of each feature were carried out using Kruskal-Wallis tests followed by post-hoc Dunn's tests [22,23]. For the during-lockdown phase, we chose the entire period of the respective national lockdown in each country, which ended when NPIs were eased for the first time. For the pre-lockdown phase, we chose the period immediately prior to the first restrictive measure with the same length of the entire respective national lockdown. For the baseline phase, we chose the same period in 2019 as the 2020 national lockdown for countries starting to collect data earlier than 2019, which included Italy, Spain, and the UK. This was aimed at suppressing seasonal variability. For Denmark and the Netherlands where participant recruitment and data collection started much later, we chose the period that started with the earliest stable date (no considerate missing data or outliers) with the same length of the entire respective national lockdown. If a significant difference among these three periods was found after Benjamini–Yekutieli correction for the number of features (9), post-hoc Dunn's test was applied with Benjamini–Yekutieli correction for the number of groups (3) [24]. Boxplots were used to present the results. A $P < .05$, after Benjamini–Yekutieli corrections, was deemed statistically significant. It should be noted that we applied corrections resulting from multiple comparisons and multiple features in each country.

We also studied factors that might influence the sub-population behavioural features during the lockdown period. The investigated factors included age, gender, body mass index (BMI), and educational background. For age groups, we defined the young group below 45 years' old and the elderly group otherwise. For BMI groups, the low BMI group was defined as below 25 and the high BMI group otherwise. For education groups, we defined the degree group as having a bachelor's degree or above and non-degree group with lower qualifications. Furthermore, we defined a combined factor group young men, as this subpopulation was suspected to be less compliant with social distancing measures. Here we focused on features of homestay and daily step count during the entire period of lockdown for each country. We performed Wilcoxon signed-rank tests on these two features to examine statistically significant differences. The $P$ were corrected with the number of factors (5) and the number of features (2) using Benjamini–Yekutieli correction.

Finally, we investigated the effects of different NPIs, in particular immediately after respective national lockdowns. This was done by comparing the NPIs implemented in the five countries within the first two weeks after entering national lockdowns.

## Results

Plots showing how the extracted features evolved from 1 February 2019 to 5 July 2020 and boxplots of these features are shown in Figure 1-5 and in Figure 6 (a-i), respectively. Detailed test statistics and $P$ values comparing pre- and during-lockdown measures are presented in Table 2. Figure 7 shows a zoomed-in version of Figure 3 and 4.

Through RADAR-base, we quantified changes in mobility, phone use, and functional measures as a result of NPIs introduced to control COVID-19. As expected, following respective national lockdowns, participants in all countries stayed at home for longer, travelled shorter distances, walked less, and made connections with fewer Bluetooth-enabled nearby devices.

In contrast to increased physical distancing (reduced sociability) suggested by these mobility features, higher phone usage suggesting compensatory sociability was observed. Italy, Spain, and the UK saw longer unlock duration, and these three countries together with the Netherlands also showed longer social app use duration. Tellingly, both unlock duration and social app use duration saw peaks around the news of national lockdowns in all countries.

Concurrent with the changes in mobility and phone usage, changes in functional measures were observed. Participants in Spain, Italy, and the UK went to bed later and slept more.

Participants in Spain, Italy, and Denmark also had a decrease in heart rate. Although not statistically significant, an increase in sleep duration and bedtime in Denmark and the Netherlands and a decrease in heart rate in the UK and Netherlands can be seen in Figure 3 to Figure 5.

The differences across countries existed in the implemented NPIs as well. The requirement of staying at home except for essential trips and the cancellation of public events were implemented in all countries but Denmark where they were only recommended. Working places were required to close for some sectors in Spain, the UK, and Denmark, and were required to close for all but essential works in the Netherlands and Italy. Public transport was recommended to close in Italy, Spain, and Denmark. Among all countries, Spain had the least strict restrictions on gatherings and school closures (only geographically targeted).

We observed that the young group spent more time at home in Italy, Spain, and the UK, and degree holders in Italy and Denmark. The young group took fewer daily steps in Italy, the UK, and the Netherlands; the low BMI group in Italy, Spain, Denmark, and the UK; and the young men group in Italy, the UK, and the Netherlands. Participants educated to degree level walked more in the UK and the Netherlands, but less in Italy. The detailed results are presented in Table 3.

**Table 2.** Results of Posthoc Dunn's test (after Kruskal-Wallis tests) on the extracted features between pre- and during-lockdown periods (only statistically significant differences were reported)

| Features | Tests | Italy | Spain | Denmark | UK | Netherlands |
|---|---|---|---|---|---|---|
| homestay | z-test statistics | -9.38 | -8.98 | -5.44 | -9.19 | -7.33 |
|  | *P-value* | <0.001 | <0.001 | <0.001 | <0.001 | <0.001 |
| Maximum distance from home | z-test statistics | 9.0 | 8.91 | 5.48 | 8.40 | 7.58 |
|  | *P-value* | <0.001 | <0.001 | <0.001 | <0.001 | <0.001 |
| Steps | z-test statistics | 8.23 | 7.72 | 2.57 | 6.82 | 4.78 |
|  | *P-value* | <0.001 | <0.001 | 0.02 | <0.001 | <0.001 |
| Maximum number of | z-test statistics | 9.68 | 8.16 | 5.06 | 10.2 | 7.73 |

| Features | Factors | Test | | | | |
|---|---|---|---|---|---|---|
| nearby devices | P-value | <0.001 | <0.001 | <0.001 | <0.001 | <0.001 |
| total sleep duration | z-test statistics | -4.65 | -5.17 | — | -4.24 | — |
| | P-value | <0.001 | <0.001 | — | <0.001 | — |
| bedtime | z-test statistics | -4.31 | -7.54 | — | -5.28 | — |
| | P-value | <0.001 | <0.001 | — | <0.001 | — |
| heart rate | z-test statistics | 9.94 | 7.61 | 2.68 | 4.18 | — |
| | P-value | <0.001 | <0.001 | 0.02 | <0.001 | — |
| unlock duration | z-test statistics | -8.8 | -8.57 | — | -6.0 | — |
| | P-value | <0.001 | <0.001 | — | <0.001 | — |
| social app use duration | z-test statistics | -7.72 | -2.36 | — | -5.72 | -4.98 |
| | P-value | <0.001 | <0.001 | — | <0.001 | <0.001 |

**Table 3.** Wilcoxon signed-rank test results on the examined factors during-lockdown periods (only statistically significant differences were reported)

| Features | Factors | Test | Italy | Spain | Denmark | UK | Netherlands |
|---|---|---|---|---|---|---|---|
| Homestay | age | w-test statistics | 108 | 252 | — | 3 | — |
| | | P-value | <0.001 | 0.003 | — | <0.001 | — |
| | gender | w-test statistics | — | — | — | — | — |
| | | P-value | — | — | — | — | — |
| | degree | w-test statistics | 270 | — | 0 | — | — |
| | | P-value | <0.001 | — | 0.004 | — | — |
| | BMI | w-test statistics | — | — | — | 220 | — |
| | | P-value | — | — | — | <0.001 | — |
| | Young men | w-test statistics | — | — | — | — | — |
| | | P-value | — | — | — | — | — |
| Step | age | w-test statistics | 283.5 | — | — | 285.5 | 151 |
| | | P-value | <0.001 | — | — | 0.004 | <0.001 |
| | gender | w-test statistics | 67.5 | 277 | — | — | — |
| | | P-value | <0.001 | 0.005 | — | — | — |
| | degree | w-test statistics | 76 | — | — | 6 | 0 |
| | | P-value | <0.001 | — | — | <0.001 | <0.001 |
| | BMI | w-test statistics | 0 | 70 | 38 | 17 | — |
| | | P-value | <0.001 | <0.001 | 0.001 | <0.001 | — |
| | Young men | w-test statistics | 149 | — | 26 | — | 0 |

| | | | | | |
|---|---|---|---|---|---|
| *P-value* | <0.001 | — | <0.001 | — | 0.02 |

## Discussion

### Principal Findings

We quantitatively investigated COVID-19 and associated lockdown related changes in mobility, functional measures, and phone usage features derived from passive data collected through mobile devices (smartphones and wearable Fitbit devices) of participants recruited in five European countries to the RADAR-CNS programme. We were able to measure significant changes in behavioural features between baseline/pre-lockdown and during-lockdown periods. As well as confirming expected changes such as spending more time at home, travelling much less, having far fewer nearby devices, we observed that people were more active on their phones, interacting with others through social apps particularly around major news events such as national lockdown, suggesting physical but maybe less social distancing. Furthermore, participants had lower heart rates, slept more, and went to bed later. In addition, we found that younger people spent more time at home and took fewer daily steps. Participants with lower BMI took more steps while maintaining comparable homestay with the higher BMI group. With 5 billion global smartphone users and 500 million smartwatch/wearable device users [25,26], we propose that the ability to generate metrics such as these is vital for evaluating NPIs efficacy.

Our mobility analyses are in line with Google mobility reports [27], where substantial reductions in mobility and increase in residential stays during respective lockdown periods were found in Italy, Spain, and the UK; Denmark and the Netherlands by comparison showed an increase in mobility trends for parks and relatively small increase in residential stays. However, in comparison to Google mobility reports which provide valuable aggregated data for short periods, RADAR-base is an open-source highly configurable platform that supports collection and analysis of participant-level mobile and phone data in near real-time with a potential for targeted interventions. Specifically, focused test and tracing may be directed to people perceived to be high risk based on their behaviour. In addition, RADAR-base was also used to collect self-reported questionnaires related to emotional well-being, functional status, and disease symptom severity of its participants [17]. Since April 2020, new questionnaires have been distributed to specifically assess COVID-19 symptoms and diagnosis status of our RADAR-CNS research participants. Our future work will use the entirety of these data to investigate the potential of wearable data such as digital early warning signs of COVID-19 and the impact of COVID-19 on the QoL and clinical trajectories of their primary diagnosis (MDD or MS).

The difference in the response across nations may reflect differences in the implementation of NPIs, media communication, and cultural differences. Denmark implemented stricter restrictions of working places and public transports but less strict on homestay and public events [21]. In contrast, Spain was more flexible on restrictions of group gatherings and school closures. The contrast in the implementation of different NPIs between the two countries showing distinct behaviours during lockdown shed light on which NPIs might be more productive in promoting physical distancing. This shows the potential utility of RADAR-base for remotely monitoring the effect of different NPIs and we also saw evidence of this in our data with participants beginning to return to pre-lockdown routines as NPIs being lifted. Future work will compare these differences within a country and across countries, which may further elaborate on the effect and impact of NPIs on infection rates and potential second wave.

It is interesting to note that the younger group in general stayed at home more and took fewer steps than the older group. Since most countries required staying at home except for essential trips, one reason could be that the elderly, often less experienced in using online shopping, had to go out for groceries. This conjecture requires future work to investigate. Those educated to degree level stayed at home for significantly longer in Italy and Denmark, possibly reflecting higher employment in sectors better able to work from home. The low BMI group took more steps but retained similar homestay to the high BMI group, which suggested they may have found other means to exercise locally. This information helps us understand the effectiveness of the NPIs at a subpopulation level and may be useful in informed strategies for targeted NPIs.

The ability to simultaneously manage multiple data modalities in RADAR-base facilitates the joint analysis and interpretation of them together with NPIs. The decrease in heart rate may be explained by the concurrent reduction in steps, the increase in homestay and total sleep duration. The reduction in mobility, coupled with an increase in phone usage, could possibly serve as indicators of physical distancing observance and resultant compensated social interaction. The delayed bedtime might be related to children homeschooling as a result of school closure, increased phone usage, and a lack of exercise. As such, RADAR-base can also be applied to monitor the population health when jointly interpreting features such as step count, sleep duration, and bedtime, which is vital if the social distancing is implemented for a longer duration.

Finally, it has been shown that an elevated resting heart rate may suggest acute respiratory infections [28]. It may be possible to infer one's infection by continuously monitoring heart rate, especially when the population remains indoor for a vast majority of the time. Such monitoring provides the possibility to generate early warning signals for symptomatic or presymptomatic respiratory infections, thereby aiding timely self-isolation or treatment. The COVID-19 related symptom and diagnosis questionnaires have been added to the study and may provide a means to investigate these relationships further.

### Media Effects

In addition to changes in trends over longer periods, we also identified interesting findings in relation to specific events (see Figure 7). A dramatic reduction in total sleep duration was observed in Denmark around 11 March 2020 which may be related to the announcement of the pending lockdown on that day and a 185% increase in the confirmed cases in Denmark on the previous day. Another example can be seen just after the mitigation phase was announced in the UK on 12 March, in which social distancing was not strongly recommended, yet we saw a trend towards participants isolating themselves voluntarily by staying at home for much longer. These observations highlight the potential role of media and social media in the distribution of information that may precipitate certain behaviour. This observation may also explain the significant difference between the baseline and pre-lockdown phases and suggests that people may have acted ahead of further government restriction. Furthermore, this is accompanied by a marked loss of weekday/weekend periodic structure pre/during the lockdown period.

**Limitations**

There are some issues to consider concerning this work. First, the participants included in this study have different medical conditions (MDD or MS), which led to different baseline levels across countries. Nevertheless, as the focus of this study is the changes in the pre-, and during-lockdown phases relative to the baseline, we were still able to identify and compare the changes induced by lockdowns. We also analysed the data collected in Spain split into MDD and MS separately. The trends and the statistical differences in all features remained the same except for total sleep duration. The sleep data from the MDD study in Spain had significant missing data, which did not meet the analysis prerequisite of 20 participants' data. Understanding of any artefacts or effects introduced into the RADAR-CNS data by the NPIs will be crucial in RADAR-CNS being able to deliver its aim of identifying signals that predict and prevent MDD and MS. Although the medical conditions of the population in this study might not be fully generalisable to a wider population healthy or with other conditions, we were able to demonstrate the utility of RADAR-base in monitoring behavioural changes, which can be readily generalised to other cohorts.

Second, the individual disease status at baseline may be different from that of the during-lockdown period in each country, which might complicate the comparisons. To mitigate this, we used the same time period from the previous year to suppress the seasonal variability. We believe this complication on the population-level was unlikely to be large, especially compared to the impact of lockdowns.

Third, participants recruited at different times may use different devices for smartphones and Fitbit depending on the availability and enrollment dates, which might make it difficult for inter-participant comparisons. Yet, this work focused on the population-level behavioural changes induced by NPIs where the handset variability was less of a factor.

Fourth, on account of requirements for participants' privacy in the RADAR-CNS studies, location data were purposely obfuscated with a participant-specific random value preventing precise localisation of the participants, which limited use of regional geographic factors within a country. It would be interesting to examine how specific regions react to lockdowns when these data are available in future work.

Fifth, limited sample sizes in certain countries and data loss impacted the smoothness of the plots showing how the extracted features evolved over time. The plots for Denmark and the Netherlands showed relatively large variance particularly in the early phase as these sites have only recently begun recruiting. Several dips and spikes in step counts and heart rate were seen in all countries during July and August. This observation was due to having data loss due to connectivity issues with the Fitbit server during this time.

Last but not least, we only explored a subset of features that can be derived from smartphones and Fitbit wearable devices. Future work will investigate whether other features offer additional information for a more complete description of lifestyle changes.

## Conclusions

Using participants' data from smartphones and wearable devices collected and managed by RADAR-base over 1.5 years covering the outbreak and subsequent spread of the COVID-19 pandemic across five European countries, we were able to detect and monitor the physical-behavioural and social-behavioural changes in response to the NPIs. We found that as well as expected findings (that validated the data collection platform) relating to increased time spent at home, less travel, and fewer nearby Bluetooth-enabled devices, participants were more active on their phone, in particular, interacting with others using social apps, particularly around major news events suggesting physical rather than social distancing. Furthermore, we found that participants had lower heart rates, slept more, and went to bed later. We demonstrated different responses across countries with Denmark showing attenuated responses to NPIs compared to other countries, which may be associated with their different focus of implementation NPIs. We found that the young stayed at home for longer yet walked less compared to the elderly and that the people with lower BMI remained more active during lockdown while having comparable homestay compared to their counterparts with higher BMI. Joint analysis of the extracted features is important for evaluating aspects of NPIs performance during their introduction and any subsequent relaxation of these measures. This work demonstrates the value of RADAR-base for collecting data from wearables and mobile technologies in order to understand the effect and response of public health interventions implemented in response to infectious outbreaks such as COVID-19. This ability to monitor response to interventions, in near real time, will be particularly important in understanding behaviour as social distancing measures are relaxed as part of any COVID-19 exit strategy. Future work will include utilising participants responses to COVID-19 related questionnaires, together with an expanded feature set to gain more specific understandings into the relationship between mobile devices derived features and the COVID-19 symptoms.

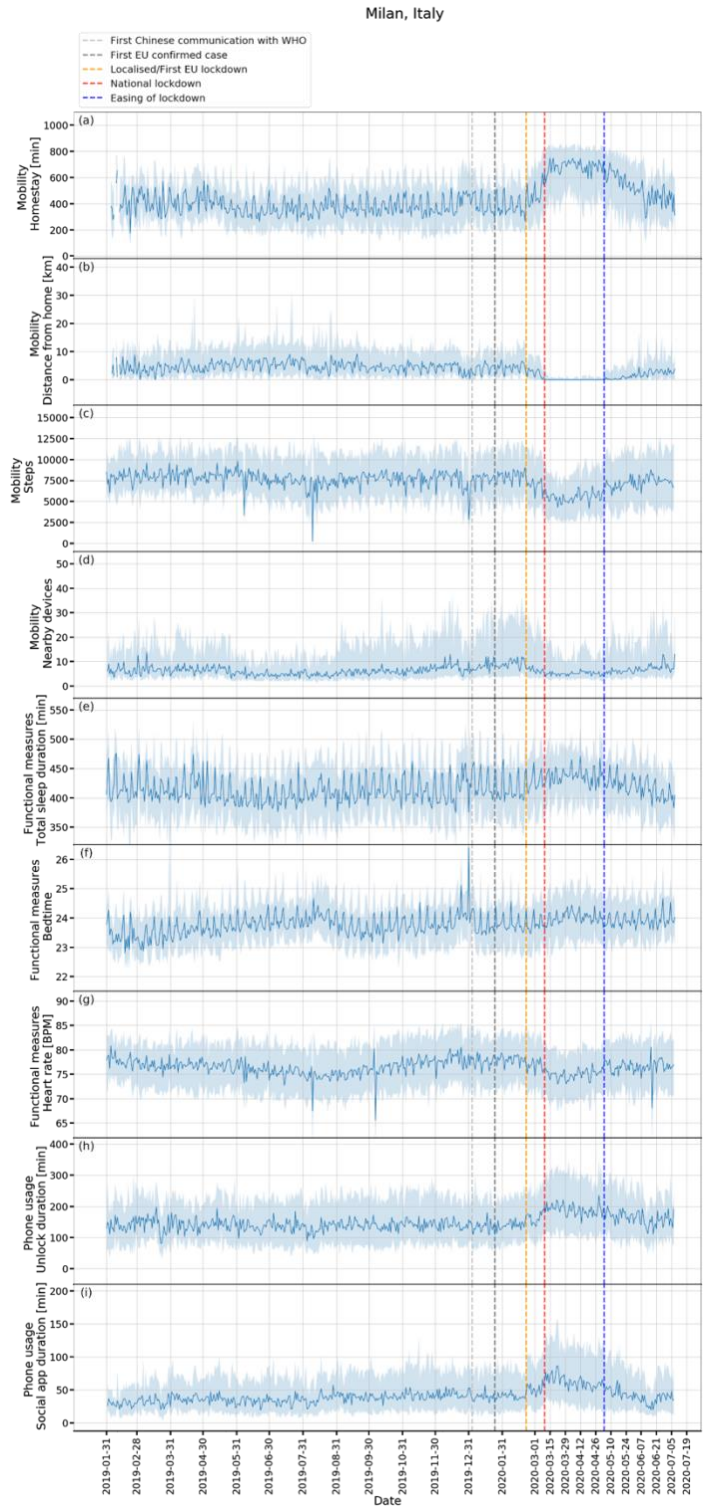

**Figure 1.** Behavioural changes for Milan, Italy (208 participants). (a): homestay duration, (b): maximum distance from home, (c): Fitbit step count, (d): maximum number of nearby devices, (e): total sleep duration, (f): bedtime, (g): heart rate, (h): unlock duration, (i): social app duration. Solid line: median, shade: 25th percentile to 75th percentile

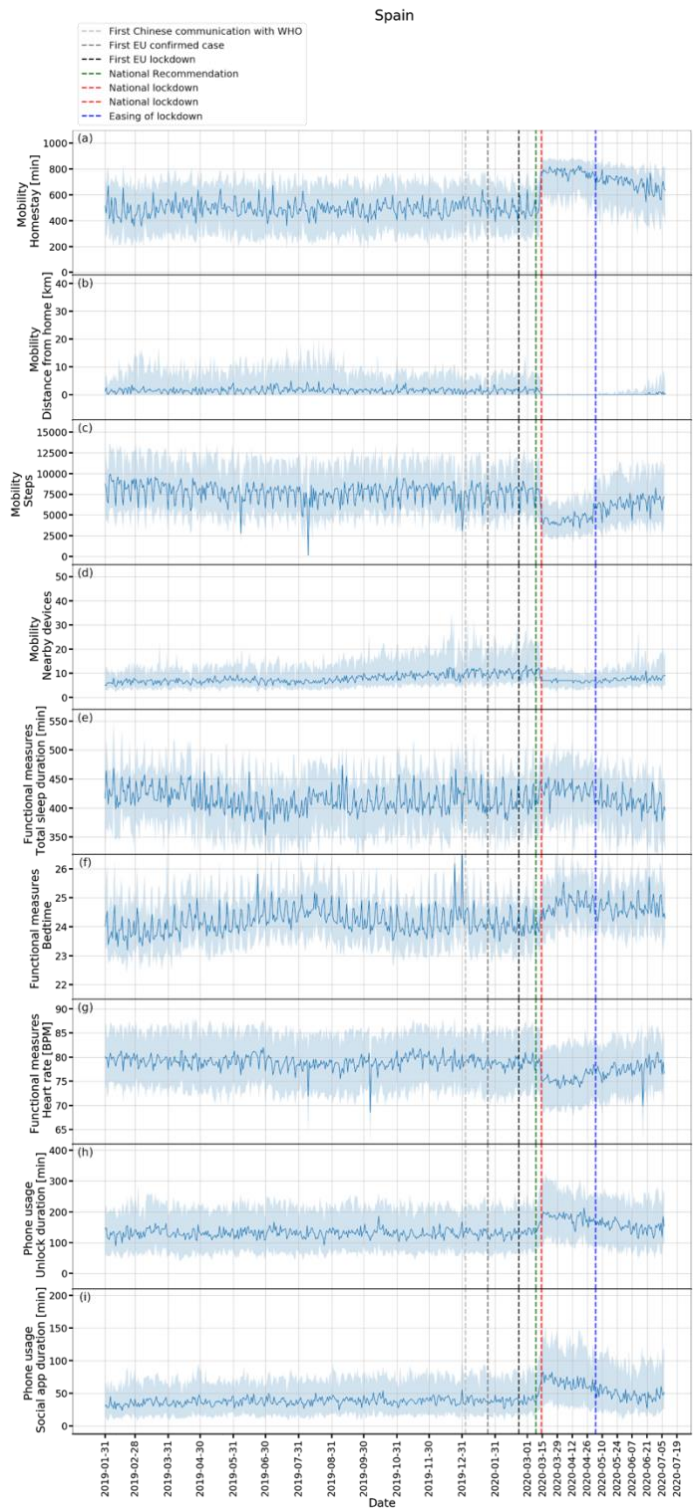

**Figure 2.** Behavioural changes for Spain (329 participants). (a): homestay duration, (b): maximum distance from home, (c): Fitbit step count, (d): maximum number of nearby devices. (e): total sleep duration, (f): bedtime, (g): heart rate, (h): unlock duration, (i): social app duration. Solid line: median, shade: 25th percentile to 75th percentile

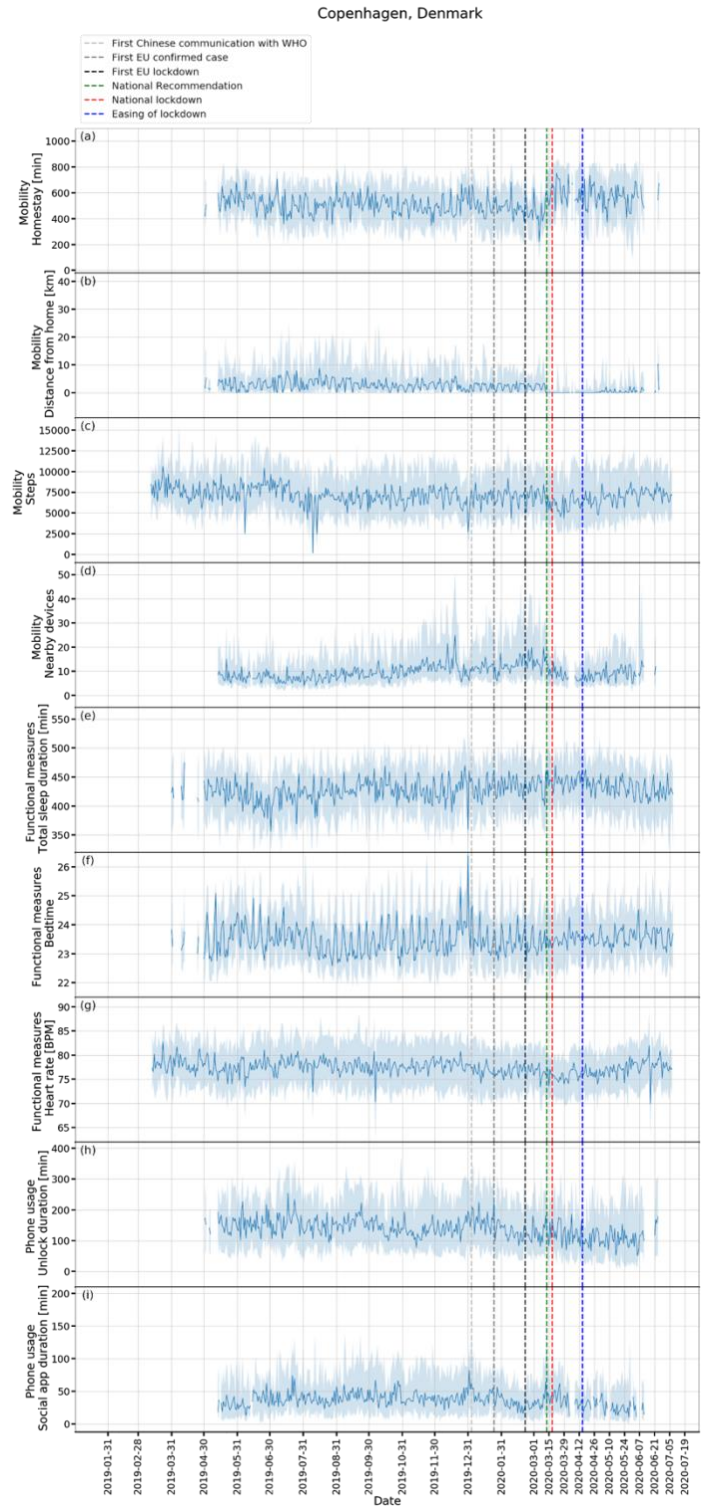

**Figure 3.** Behavioural changes for Copenhagen, Denmark (106 participants). (a): homestay duration, (b): maximum distance from home, (c): Fitbit step count, (d): maximum number of nearby devices. (e): total sleep duration, (f): bedtime, (g): heart rate, (h): unlock duration, (i): social app duration. Solid line: median, shade: 25th percentile to 75th percentile

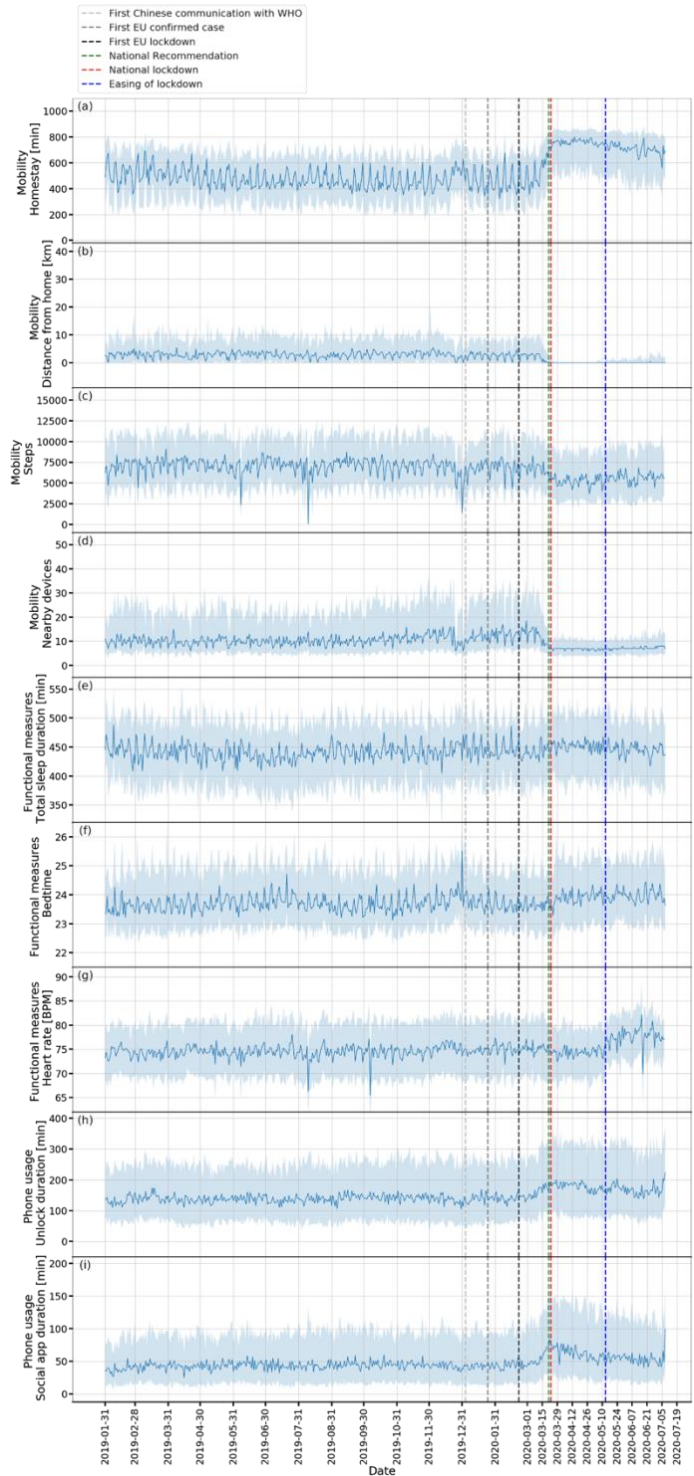

**Figure 4.** Behavioural changes for London, the United Kingdom (316 participants). (a): homestay duration, (b): maximum distance from home, (c): Fitbit step count, (d): maximum number of nearby devices. (e): total sleep duration, (f): bedtime, (g): heart rate, (h): unlock duration, (i): social app duration. Solid line: median, shade: 25th percentile to 75th percentile

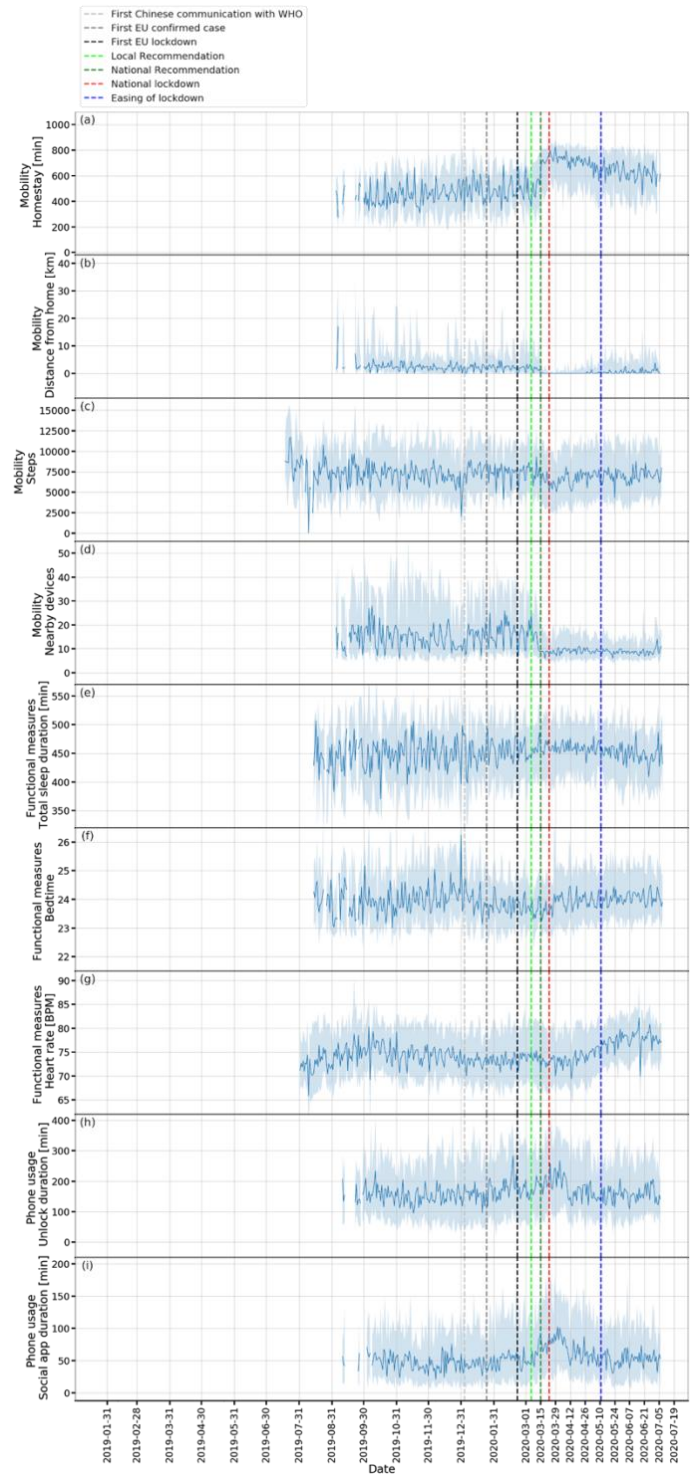

**Figure 5.** Behavioural changes for Amsterdam, the Netherlands (103 participants). (a): homestay duration, (b): maximum distance from home, (c): Fitbit step count, (d): maximum number of nearby devices. (e): total sleep duration, (f): bedtime, (g): heart rate, (h): unlock duration, (i): social app duration. Solid line: median, shade: 25th percentile to 75th percentile

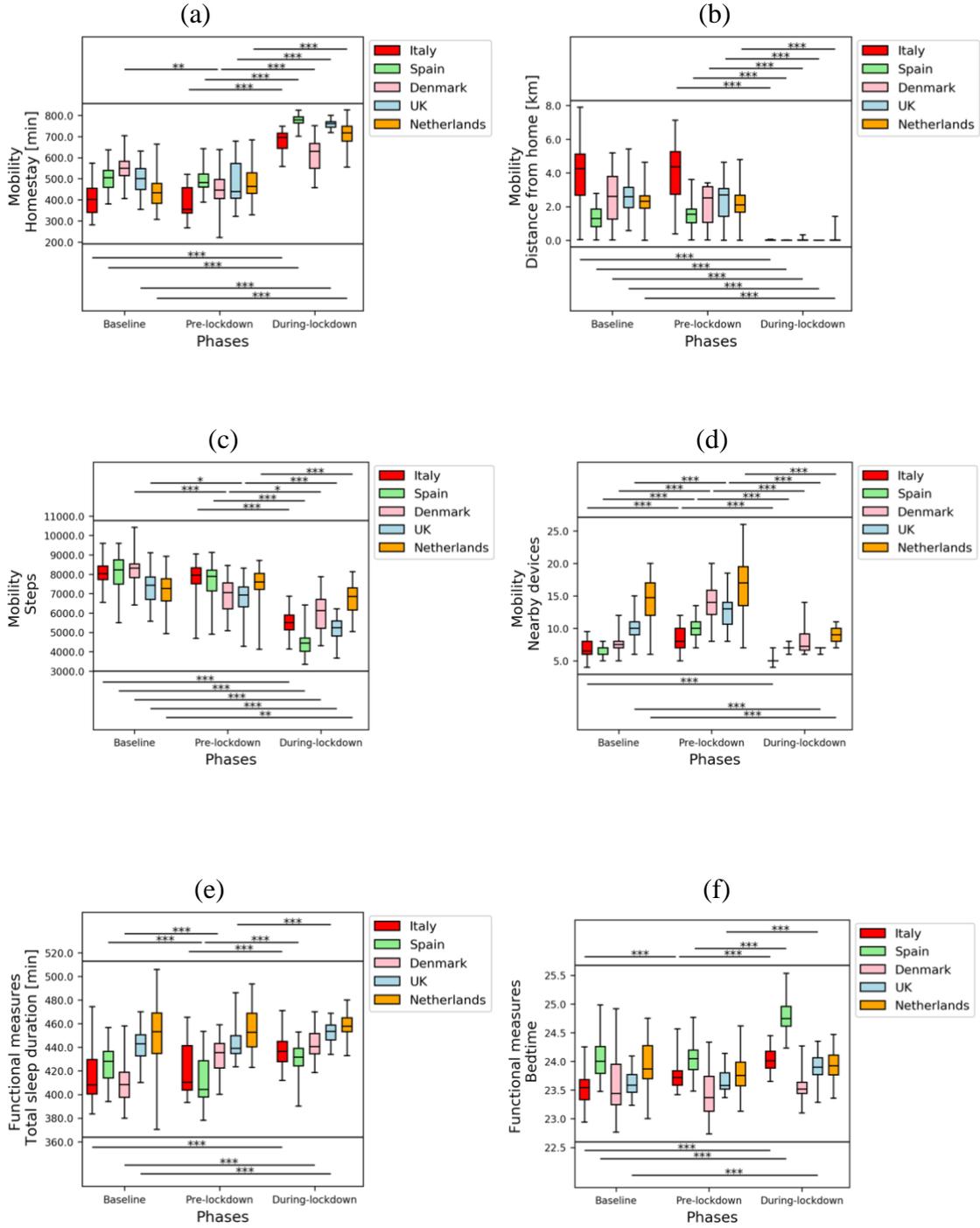

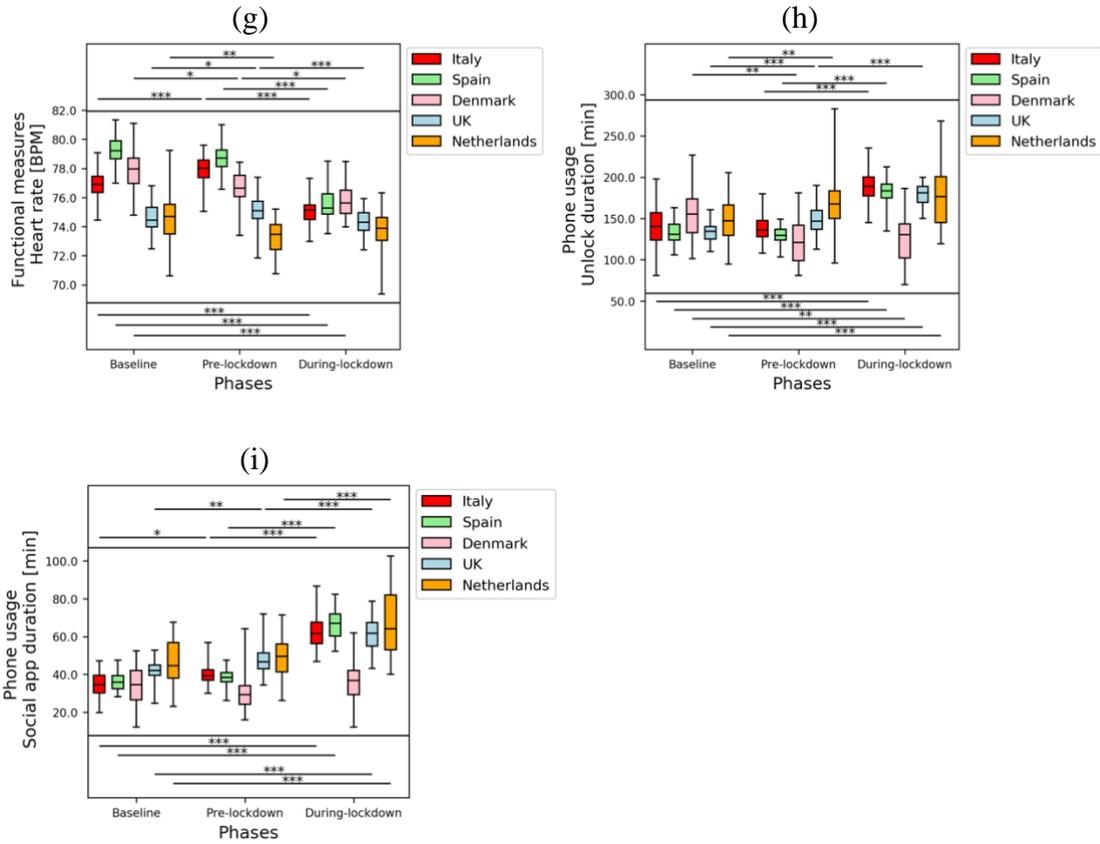

**Figure 6.** Boxplots for comparisons among baseline, pre- and during-lockdown phases for different features. * means $p < 0.05$, ** means $p < 0.01$, ** means $p < 0.001$. (a): homestay duration, (b): maximum distance from home, (c): Fitbit step count, (d): maximum number of nearby devices. (e): total sleep duration, (f): bedtime, (g): heart rate, (h): unlock duration, (i): social app duration.

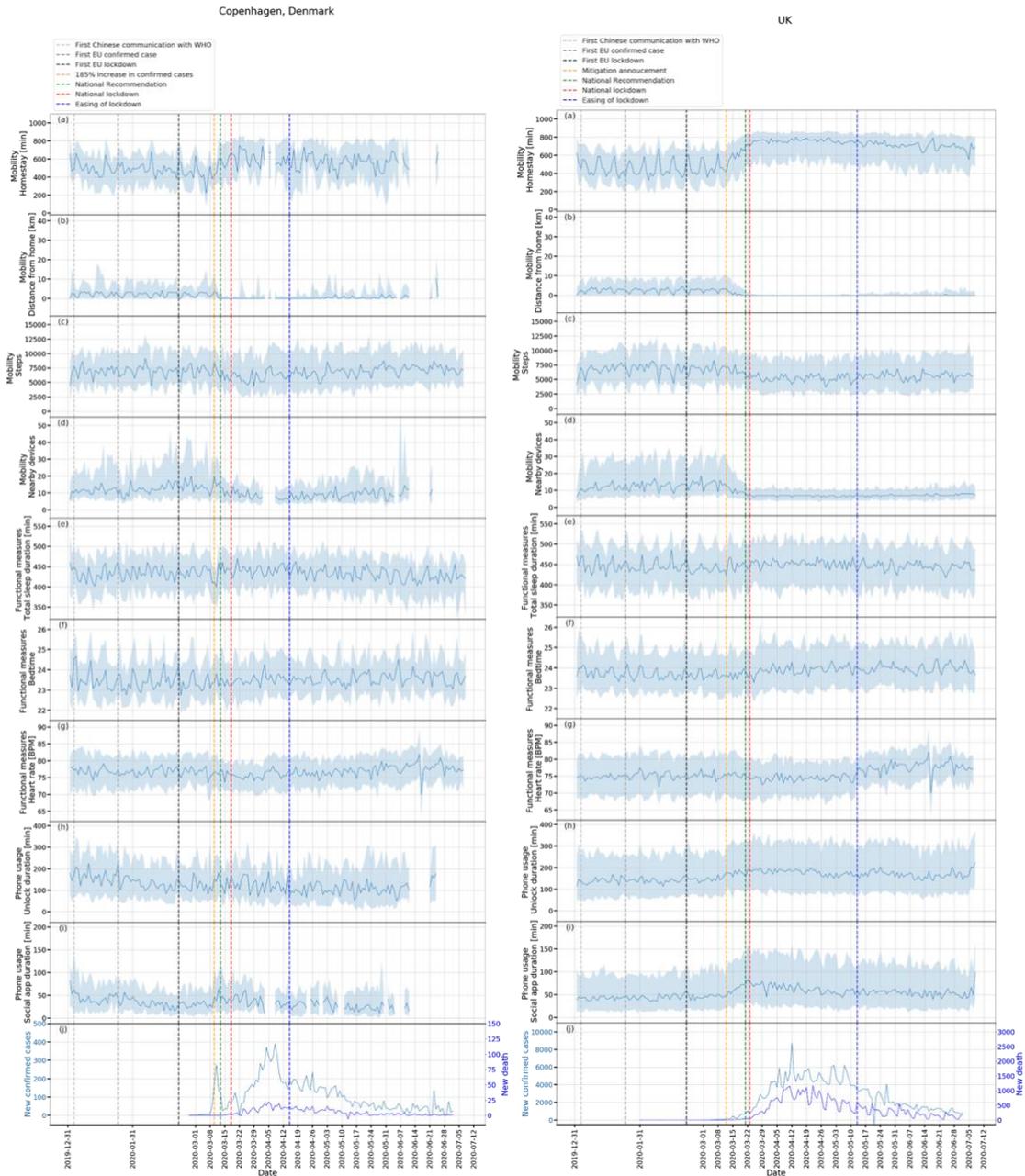

**Figure 7.** Zoomed-in plots for Copenhagen, Denmark and the UK. (a): homestay duration, (b): maximum distance from home, (c): Fitbit step count, (d): maximum number of nearby devices. (e): total sleep duration, (f): bedtime, (g): heart rate, (h): unlock duration, (i): social app duration. (j): COVID-19 confirmed and death cases. Solid line: median, shade: 25th percentile to 75th percentile

**Acknowledgements**

This study was supported by National Institute for Health Research (NIHR) Biomedical Research Centre at South London and Maudsley NHS Foundation Trust, King's College London, and EU/EFPIA IMI Joint Undertaking 2 (RADAR-CNS grant No 115902). This communication reflects the views of the RADAR-CNS consortium, and neither IMI nor the European Union and EFPIA are liable for any use that may be made of the information contained herein. Participant recruitment in Amsterdam, the Netherlands was partially accomplished through Hersenonderzoek.nl, the Dutch online registry that facilitates participant recruitment for neuroscience studies (www.hersenonderzoek.nl). Hersenonderzoek.nl is funded by ZonMw-Memorabel (project no. 73305095003), a project in the context of the Dutch Deltaplan Dementie, Gieskes-Strijbis Foundation, the Alzheimer's Society in the Netherlands (Alzheimer Nederland) and Brain Foundation Netherlands (Hersenstichting). This study has also received support from Health Data Research UK (funded by the UK Medical Research Council), Engineering and Physical Sciences Research Council, Economic and Social Research Council, Department of Health and Social Care (England), Chief Scientist Office of the Scottish Government Health and Social Care Directorates, Health and Social Care Research and Development Division (Welsh Government), Public Health Agency (Northern Ireland), British Heart Foundation and Wellcome Trust, and The National Institute for Health Research University College London Hospitals Biomedical Research Centre.

## Authors' Contributions

SS, AAF, and RJBD contributed to the study design. SS contributed to the data analysis, Figures drawing, and manuscript writing. AAF, NC, TW, VAN, GC, MH and RJBD contributed to the critical revision of the manuscript. AAF, YR, ZR, PC, CS, and RJBD contributed to the platform design and implementation. AAF, IMG, AR, TW, VAN, GC, MH, and RJBD contributed to the administrative, technical and clinical support of the study. FM, GDC, SS, LL, ALG, AZ, BWJHP, FL, SS, JMH contributed to data collection.

## Conflicts of Interest

VAN is an employee of Janssen Research & Development LLC and may own equity in the company.

## Abbreviations

NPIs: Non-Pharmaceutical Interventions
WHO: World Health Organisation
GPS: Global Positioning System
MDD: Major Depressive Disorder
MS: Multiple Sclerosis
QoL: Quality of Life